\documentclass[cite]{epl}

\def\lambdabar {\mathchar'26\mkern-10mu\lambda}

\title{On a dynamical symmetry group of the relativistic linear singular oscillator}
\shorttitle{On a dynamical symmetry group of the relativistic ...}
\author{S.M. Nagiyev\inst{1}, E.I. Jafarov\inst{1}\thanks{E-mail: \email{azhep@physics.ab.az}} \and R.M. Imanov\inst{2}}
\shortauthor{S.M. Nagiyev \etal}
\institute{                    
  \inst{1} Institute of Physics, Azerbaijan National Academy of Sciences - Javid ave. 33, AZ1143, Baku, Azerbaijan\\
  \inst{2} Physics Department, Ganja State University - A. Camil str. 1, AZ2700, Ganja, Azerbaijan
}

\begin{document}

\maketitle

\begin{abstract}
An exact approach for the factorization of the relativistic linear singular oscillator is proposed. This model is expressed by the finite-difference Schr\"odinger-like equation. We have found finite-difference raising and lowering operators, which are with the Hamiltonian operator form the close Lie algebra of the $SU\left( 1, 1 \right)$ group.
\end{abstract}

\section{Introduction}
\label{int}

The singular harmonic oscillator is one of the rare exactly solvable problems in non-relativistic quantum mechanics \cite{landau,calogero}. This model is useful to explain many phenomena, such as description of interacting many-body systems \cite{dodonov}, diatomic \cite{chumakov} and polyatomic \cite{hartmann} molecules, spin chains \cite{polychronakos}, quantum Hall effect \cite{frahm}, fractional statistics and anyons \cite{leinaas}. In spite of many interesting papers devoted to the study of the non-relativistic singular harmonic oscillator model \cite{maamche}, the number of works studying relativistic approachs to singular oscillator exact solution is still rather few \cite{castro}.

Recently, we constructed the exactly solvable relativistic model of the quantum linear singular oscillator \cite{nagiyev}. Later, this model was generalized for 3D case in \cite{nagiyev2}. Our model was formulated in the framework of the finite-difference version of the relativistic quantum mechanics, developed in \cite{kadyshevsky1,freeman,klein,atakishiyev1,atakishiyev2}. As we noted in \cite{nagiyev}, unlike the case of the Coulomb potential, the relativistic generalization of the oscillator or singular oscillator potentials are not uniquely defined. Therefore, one of the main requirements for proposed relativistic models is existence of their dynamical symmetry. In this Letter, we present the simplest way for construction of the close Lie algebra of the $SU \left(1, 1\right)$ group for a relativistic model of the linear singular oscillator. In spite of expression of the problem under consideration by the finite-difference equation, it is also exactly factorizable and is in complete analogy with relevant non-relativistic problem. This factorization will allow us to obtain the exact finite-difference expression of the generators of $SU \left(1, 1\right)$ group.

The Letter is structured by following: Section 2 is devoted to brief description of the finite-difference relativistic quantum mechanics, whereas the Section 3 describes this approach in 1D. We do some brief review of the factorization scheme for the non-relativistic linear singular oscillator in Section 4 and present the completely similar factorization scheme for a relativistic model of the linear singular oscillator in Section 5.

\section{Finite-difference formulation of the relativistic quantum mechanics}

In the finite-difference relativistic quantum mechanics the relative motion wavefunction satisfies a finite-difference equation with a step equal to the Compton wavelength of the particle, $\lambdabar=\hbar/mc$. For example, in the case of a local quasipotential of interaction $V\left( \vec r \right) $ the equation for the wavefunction of two scalar particles with equal mass has the form

\begin{equation}
\label{1}
\left[ H_0+V\left( \vec r \right) \right] \psi (\vec r)=E\psi (\vec r) \; ,
\end{equation}
with the free relativistic Hamiltonian operator

\begin{equation}
\label{2}
H_0=mc^2\left[ \cosh \left( i\lambdabar \partial _r\right) + \frac{i \lambdabar}r\sinh \left( i \lambdabar \partial _r\right) + \frac{  {\vec L}^2}{2\left( mcr \right) ^2}\exp \left( i\lambdabar \partial _r\right) \right]  \; ,
\end{equation}
where $\vec L$ is the orbital angular momentum operator and $\partial _r  \equiv \frac{\partial }{{\partial _r }}$. The technique of difference differentiation was developed and analogues of the important functions of the continuous analysis were obtained to fit the relativistic quantum mechanics, based on Eq.(\ref{1}) \cite{kadyshevsky1}.

The space of vectors $\vec r$ is called the relativistic configurational space or $\vec r$-space. The transformation between $\vec r$-space and its canonically conjugate momentum $\vec p$-space is given by the 'relativistic plane waves' \cite{kadyshevsky1,shapiro}

\begin{eqnarray}
\label{3}
\xi \left( \vec p,\vec r\right) =\left( \frac{p_0-\vec p\vec n}{mc}\right) ^{-1-ir/\lambdabar } \; ,  \\
\vec r=r\vec n, \; {\vec n}^2=1, \quad 0\leq r<\infty ,\nonumber \\ p_0=\sqrt{{\vec p}^2+m^2c^2} \nonumber
\end{eqnarray}
rather than by the usual one $\exp \left(i \vec p \vec r / \hbar \right)$ in the non-relativistic case. These functions correspond to the principal series of unitary representations of the Lorentz group $SO\left(3, 1 \right)$ and in the non-relativistic limit ($c \to \infty$) we have $\xi \left( \vec p,\vec r\right) \to \exp \left(i \vec p \vec r / \hbar \right)$. The momenta of particles belong to the three dimensional Lobachevsky space realized on the upper sheet of the mass hyperboloid ${p_0}^2-{\vec p}^2=m^2c^2$.

The relativistic plane waves (\ref{3}) are the eigenfunctions of the free Hamiltonian operator (\ref{2}), i.e.

\begin{equation}
\label{4}
\left(H_0-E_p \right) \xi \left(\vec p, \vec r \right) = 0, \quad E_p  = cp_0  = c\sqrt {\vec p^2  + m^2 c^2 }.
\end{equation}

They form a complete and orthogonal system of functions, i.e. 

\begin{eqnarray}
\label{5}
 \frac{1}{{\left( {2\pi \hbar } \right)^3 }}\int {\xi ^* \left( {\vec p,\vec r} \right)\xi \left( {\vec p,\vec r'} \right)d\Omega _p }  = \delta \left( {\vec r - \vec r'} \right), \\ 
 \frac{1}{{\left( {2\pi \hbar } \right)^3 }}\int {\xi ^* \left( {\vec p,\vec r} \right)\xi \left( {\vec p',\vec r} \right)d\vec r}  = \delta \left( {\vec p\left(  -  \right)\vec p'} \right) = \frac{{p_0 }}{{mc}}\delta \left( {\vec p - \vec p'} \right),
\end{eqnarray}
where $d\Omega _p  = mc\frac{{d\vec p}}{{p_0 }}$ is the Lorentz invariant volume element of the Lobachevsky space.

\section{1D Relativistic Quantum Mechanics}

In the one-dimensional case the relativistic plane wave takes a form \cite{atakishiyev1}

\begin{equation}
\label{6}
\xi \left( p,x\right) =\left( \frac{p_0-p}{mc}\right) ^{-ix/\lambdabar } \; ,
\end{equation}
or, in hyperpolar coordinates

\begin{equation}
\label{7}
p_0=mc \cosh \chi \; , \quad p=mc \sinh \chi \; 
\end{equation}
we have the exponential function

\begin{equation}
\label{8}
\xi (p,x) = e^{ix \chi / \lambdabar } \; , 
\end{equation}
where $\chi= \ln \left( \frac{p_0+p}{mc} \right)$ is rapidity.

The functions (\ref{6}) obey the free relativistic finite-difference Schr\"odinger equation

\begin{equation}
\label{9}
\left( {\hat H_0  - E_p } \right)\xi \left( {p,x} \right) = 0,
\end{equation}
with the simple free Hamiltonian operator $\hat H_0  = mc^2 \cosh \left( {i\lambdabar \partial _x } \right)$.

Note that all the important exactly solvable cases of non-relativistic quantum mechanics (potential well, Coulomb potential, harmonic oscillator, singular oscillator etc.) are also exactly solvable for the case of equation (\ref{1}) (or for the corresponding one-dimensional equation). In this connection one can note, that there is a well-studied class of the relativistic one-dimensional integrable quantum many-body systems \cite{ruijsenaars}. They provide a relativistic generalization of the Calogero-Sutherland systems \cite{calogero}. In principle, the approach used in this Letter can reproduce the Ruijsenaars-Schneider models. We could not find any relevant paper studying comparatively quasipotential and Ruijsenaars-Schneider finite-difference generalizations of the well-known non-relativistic quantum mechanics. Therefore, the paper devoted to this problem now is in preparation process.

\section{Factorization of the non-relativistic linear singular harmonic oscillator}

In the non-relativistic quantum mechanics the Hamiltonian of the linear singular harmonic oscillator

\begin{equation}
\label{11}
H_N  =  - \frac{{\hbar ^2 }}{{2m}}\partial _x ^2  + \frac{{m\omega ^2 }}{2}x^2  + \frac{g}{{x^2 }}
\end{equation}
may be factorized as follows:

\begin{equation}
\label{12}
H_N  = \hbar \omega \left( {c^ +  c^ -   + d + 1} \right)
\end{equation}
in terms of the operators, having the form (see, for example, \cite{nagiyev})

\begin{equation}
\label{13}
c^ \pm   = \frac{1}{{\sqrt 2 }}\left( {\xi  \mp \partial _\xi   - \frac{{d + 1/2}}{\xi }} \right) = a^ \pm   - \frac{{d + 1/2}}{{\sqrt 2 \xi }},\;\xi  = \sqrt {\frac{{m\omega }}{\hbar }} x.
\end{equation}

Here $a ^ \pm$ are the usual creation and annihilation operators and $d = \frac{1}{2}\sqrt {1 + \frac{{8mg}}{{\hbar ^2 }}} $. We suppose $g> - \hbar ^2/8m$, so $d$ is a real positive number \cite{dodonov}.

Using $c^-$ and $c^+$, one can construct a dynamical symmetry algebra of the linear singular oscillator (\ref{11}). For this purpose we calculate their commutator:

\begin{equation}
\label{14}
\left[ {c^ -  ,c^ +  } \right] = 1 + \frac{{d + 1/2}}{{\xi ^2 }}.
\end{equation}

The right-hand side of (\ref{14}) requires the introduction of a new operator $ \xi c^-$ that leads to commutator

\begin{equation}
\label{15}
\left[ {H_N ,\xi c^ -  } \right] =  - 2\hbar \omega \left( {\xi c^ -   - \frac{{H_N }}{{\sqrt 2 \hbar \omega }} + \frac{{d + 1}}{{\sqrt 2 }}} \right),
\end{equation}
allowing us to determine exact expression of the lowering operator as follows:

\begin{equation}
\label{16}
A_N ^ -   = \sqrt 2 \xi c^ -   - \frac{{H_N }}{{\hbar \omega }} + d + 1=\left(a^- \right)^2 - \frac{g_0}{\xi^2}.
\end{equation}

Indeed, from (\ref{15}) and (\ref{16}) follows that

\begin{equation}
\label{17}
\left[ {H_N ,A_N ^ -  } \right] =  - 2\hbar \omega A_N ^ -  .
\end{equation}

Corresponding raising operator $A_N ^+=\left(a^+ \right)^2 - \frac{g_0}{\xi^2}$ can be determined as the Hermitian conjugate of the (\ref{16}).

%\begin{equation}
%\label{18}
%\left[ {H_N ,A_N ^ +  } \right] =  2\hbar \omega A_N ^ +  .
%\end{equation}

%We can express $A_N^\pm$ in terms of the creation and annihilation %operators as follows \cite{calogero}

%\begin{equation}
%\label{19}
%A_N ^ -   = \left( {a^ -  } \right)^2  - \frac{{g_0 }}{{\xi ^2 %}},\quad A_N ^ +   = \left( {a^ +  } \right)^2  - \frac{{g_0 }}{{\xi %^2 }}.
%\end{equation}

%To obtain the closed algebra of operators $A_N ^-$, $A_N^ +$ and $H_N$ %in addition to (\ref{17}) and (\ref{18}) it is necessary to calculate %following commutator:

%\begin{equation}
%\label{20}
%[ A_N^-, A_N^+]=\frac 4{\hbar \omega} H_N.
%\end{equation}

The operators $K^-=\frac 12 A_N^-$ and $K^+=\frac 12 A_N^+$ together with $K_0=H_N/2$ satisfy the canonical commutation relations

\begin{equation}
\label{21}
\left[ {K_0 ,K^ \pm  } \right] =  \pm K^ \pm  ,\;\left[ {K^ -  ,K^ +  } \right] = 2K_0 ,
\end{equation}
which define the Lie algebra of the group $SU(1,1)$.

A direct calculation shows that the invariant Casimir operator in this case is

\begin{equation}
\label{22}
K^2  = K_0 \left( {K_0  - 1} \right) - K^ +  K^ -   = \frac{{d + 1}}{2}\left( {\frac{{d + 1}}{2} - 1} \right).
\end{equation}

The eigenvalue $\frac{{d + 1}}{2}$ of the Casimir operator $K^2$ determines that the non-relativistic linear singular oscillator (\ref{11}) realizes the unitary irreducible representation $D^+ \left(\frac{{d + 1}}{2} \right)$ of the $SU(1,1)$ group. The eigenvalues of the compact generator $K_0$ in such representations are restricted from below and equal to $\frac{{d + 1}}{2}+n$, $n=0,1,2,3, \dots$. Thus, a purely algebraic approach enables one to find the correct spectrum of the Hamiltonian $H_N= 2 \hbar \omega K_0$ (\ref{11}):

\begin{equation}
\label{23}
E_n^{non-rel}= \hbar \omega \left( 2 d +n+1 \right).
\end{equation}

Due to commutation relations (\ref{21}) the action of the raising and lowering operators $K^+$ and $K^-$ on the eigenfunctions $\psi_n^{\left( N \right)}$ of the $H_N$ (\ref{11}) is given by

\begin{equation}
\label{24}
K^ +  \psi _n ^{\left( N \right)}  = \kappa _{n + 1} \psi _{n + 1} ^{\left( N \right)} ,\;K^ -  \psi _n ^{\left( N \right)}  = \kappa _n \psi _{n - 1} ^{\left( N \right)} ,
\end{equation}
where $\kappa _n  = \sqrt {n\left( {n + d} \right)}$. Hence the functions $\psi _n ^{\left( N \right)}$ can be obtained by $n$-fold application of the operator $K^+$ to the ground state wavefunction $\psi _0 ^{\left( N \right)}$, i.e.

\begin{eqnarray}
\label{25}
\psi _{\rm n} ^{\left( N \right)}  = \gamma _n \left( {K^ +  } \right)^n \psi _{\rm 0} ^{\left( N \right)}=c_n^{(N)} \xi ^{d+\frac {1}{2}} e^{- \frac{\xi^2}{2}}L_n^d \left( \xi^2 \right) , \\ 
\gamma _n ^{ - 1}  = \kappa _1 \kappa _2  \ldots \kappa _n  = \sqrt {n!\left( {d + 1} \right)_n } , \nonumber
\end{eqnarray}
where $L _n^d \left( y \right)$ are the associated Laguerre polynomials and $\left(a \right)_n = a(a + 1) \ldots (a + n - 1) = \Gamma (a + n)/\Gamma (a)$ is the Pochhammer symbol.

\section{Factorization of the relativistic linear singular oscillator}

The relativistic linear singular oscillator considered in \cite{nagiyev} is described by Hamiltonian

\begin{equation}
\label{26}
H = mc^2 \cosh \left( {i\lambdabar \partial _x } \right) + \frac{1}{2}m\omega ^2 x\left( {x + i \lambdabar } \right)e^{i\lambdabar \partial _x }  + \frac{g}{{x\left( {x + i \lambdabar } \right)}}e^{i\lambdabar \partial _x } .
\end{equation}

In terms of the dimensionless variable $\rho  = x/\lambdabar $ and parameters $\omega _0  = \frac{{\hbar \omega }}{{mc^2 }}$ and $g_0  = \frac{{mg}}{{\hbar ^2 }}$ the operator (\ref{26}) takes the form

\begin{equation}
\label{27}
H = mc^2 \left[ {\cosh \left( {i\partial _\rho  } \right) + \frac{1}{2}\omega _0 ^2 \rho ^{(2)} e^{i\partial _\rho  }  + \frac{{g_0 }}{{\rho ^{(2)} }}e^{i\partial _\rho  } } \right],
\end{equation}
where $\rho ^{(\lambda)}=i^{\left( \lambda \right)} \Gamma \left( \lambda - i \rho \right)/\Gamma \left( - i \rho \right)$ is a generalized degree \cite{freeman}.

As in the non-relativistic case (\ref{11}), it is possible to factorize \cite{nagiyev} the Hamiltonian (\ref{27})

\begin{equation}
\label{28}
H = mc^2 \left[ {b^ +  b^ -   + \omega _0 \left( {\alpha  + \nu } \right)} \right]
\end{equation}
by means of the difference operators

\begin{eqnarray}
\label{29}
b^-= \frac {1}{\sqrt{2}} \left[e ^ {-\frac {i}{2} \partial _\rho} - \omega _0 e ^ {\frac {i}{2} \partial _\rho} \left( \nu +i \rho\right) (1+ \frac {\alpha}{i \rho}) \right] \;, \\
b^+= \frac {1}{\sqrt{2}} \left[e ^ {-\frac i2 \partial _\rho} - \omega _0  \left( \nu -i \rho \right) (1- \frac {\alpha}{i \rho}) e ^ {\frac i2 \partial _\rho} \right] \; \nonumber.
\end{eqnarray}
with the notations

\begin{eqnarray}
\label{30}
\alpha =\frac 12+\frac 12\sqrt{1+\frac 2{\omega _0^2}\left( 1-\sqrt{1-8g_0\omega _0^2}\right) } \; , \\
\nu =\frac 12+\frac 12\sqrt{1+\frac 2{\omega _0^2}\left( 1+\sqrt{1-8g_0\omega _0^2}\right) } \; \nonumber.
\end{eqnarray}

To obtain a dynamical symmetry algebra of the relativistic linear singular oscillator (\ref{27}), we consider a commutator of $b^-$ and $b^+$

\begin{eqnarray}
\label{31}
 \left[ {b^ -  ,b^ +  } \right] = \frac{1}{2}\omega _0 \left( {1 + \frac{{\alpha \nu }}{{\rho ^2  + \frac{1}{4}}} + \omega _0 \Delta e^{i\partial _\rho  } } \right), \\ 
 \Delta  = \alpha  + \nu  - \frac{1}{4} + \alpha \nu \left[ { - \frac{{\left( {\alpha  - 1} \right)\left( {\nu  - 1} \right)}}{{\rho ^{(2)} }} + \frac{{\alpha \nu }}{{\left( {\rho  + i/2} \right)^{(2)} }}} \right]. \nonumber 
\end{eqnarray}

Using (\ref{31}), now one can determine raising and lowering operators through $b^+$ and $b^-$. Indeed, it is easy to show that the expression

\begin{equation}
\label{32}
B^ -   = i\rho \left[ {\sqrt 2 e^{ - \frac{i}{2}\partial _\rho  } b^ -   - \frac{H}{{mc^2 }} + \omega _0 \left( {\alpha  + \nu } \right)} \right] + \frac{1}{2}\frac{H}{{mc^2 }} - \frac{1}{{2\omega _0 }}\frac{{H^2 }}{{\left( {mc^2 } \right)^2 }} + \omega _0 \alpha \nu \;
\end{equation}
is the lowering operator, i.e.

\begin{equation}
\label{33}
\left[ {H,B^ -  } \right] =  - 2\hbar \omega B^ -  ,
\end{equation}
moreover, it is obvious the analogue with the (\ref{16}) case. The raising operator $B^+$ can be obtained as Hermitian conjugate of (\ref{32}).

By the determination of the generalized momentum operator $P$ form 
\begin{equation}
\label{34}
P =  - mc\left[ {\sinh \left( {i\partial _\rho  } \right) + \frac{1}{2}\omega _0 ^2 \rho ^{(2)} e^{i\partial _\rho  }  + \frac{{g_0 }}{{\rho ^{(2)} }}e^{i\partial _\rho  } } \right]
\end{equation}
through commutation relation

\begin{equation}
\label{35}
\left[ \rho, H \right] = i c P,
\end{equation}
one can express $B^+$ and $B^-$ in compact form

%\begin{eqnarray}
%\label{36}
% B^ -   = \frac{1}{{2\omega _0 }}\left[ {\left( {\omega _0 \rho  + %\frac{i}{{mc}}P} \right)^2  - \frac{{2g_0 }}{{\rho ^2  + 1}}} \right], %\\ 
% B^ +   = \frac{1}{{2\omega _0 }}\left[ {\left( {\omega _0 \rho  - %\frac{i}{{mc}}P} \right)^2  - \frac{{2g_0 }}{{\rho ^2  + 1}}} \right]. %\nonumber 
% \end{eqnarray}
\begin{equation}
\label{36}
 B^ \pm   = \frac{1}{{2\omega _0 }}\left[ {\left( {\omega _0 \rho  \mp \frac{i}{{mc}}P} \right)^2  - \frac{{2g_0 }}{{\rho ^2  + 1}}} \right].
\end{equation}

To obtain the closed algebra of operators $B^-$, $B^+$ and $H$ besides (\ref{33}) it is necessary to calculate the commutator

\begin{equation}
\label{37}
\left[ {B^ -  ,B^ +  } \right] = \frac{{\omega _0 }}{{mc^2 }}H\left\{ {1 + \frac{2}{{\omega _0 ^2 }}\left[ {\left( {\frac{H}{{mc^2 }}} \right)^2  - 1} \right]} \right\}.
\end{equation}

As follows from (\ref{37}), to construct the algebra of dynamical symmetry of the model under consideration it is sufficient to introduce following operators:

\begin{equation}
\label{38}
K^ -   = B^ -  f^{ - 1/2} \left( H \right),\;K^ +   = f^{ - 1/2} \left( H \right)B^ +  ,
\end{equation}
where $f\left( H \right) = \left[ {\frac{H}{{mc^2 }} + \omega _0 \left( {\alpha  - \nu  - 1} \right)} \right]\left[ {\frac{H}{{mc^2 }} + \omega _0 \left( {\nu  - \alpha  - 1} \right)} \right]$.

Now one can see that $K^ -$, $K^ +$ and $K_0  = \frac{1}{{2\hbar \omega }}H$ satisfy commutation relations (\ref{21}) on the basis elements $\left\{ {\psi _n } \right\}_{n = 0}^\infty  $, i.e.

$$
\left[ {K^ -  ,K^ +  } \right]\psi _n  = 2K_0 \psi _n,
$$
where $\psi _n$ are eigenfunctions of $H$ \cite{nagiyev}. Their explicit form will be provided again below. Moreover the Casimir operator in this case is equal to $K^2  = \frac{{\alpha  + \nu }}{2}\left( {\frac{{\alpha  + \nu }}{2} - 1} \right)$.

Hence, eigenvalues of $H=2\hbar \omega K_0$ are restricted from below and are equal to $E_n=\hbar \omega \left(2n+ \alpha + \nu \right)$, $n=0,1,2, \dots$, and eigenfunctions $\psi _n$ form the basis of the irreducible representation $D^ +  \left( {\frac{{\alpha  + \nu }}{2}} \right)$ of the $SU\left( 1,1 \right)$ group.

In conclusion one can note that action of $B^+$ and $B^-$ operators to eigenfunctions $\psi _n$ of the Hamiltonian (\ref{26}) can be expressed by formulae

\begin{eqnarray}
\label{40}
 B^ -  \psi _n  = b_n \psi _{n - 1} ,\;B^ +  \psi _n  = b_{n + 1} \psi _{n + 1} , \\ 
 b_n  = 2\omega _0 \sqrt {n\left( {n + \alpha  + \nu } \right)\left( {n + \alpha  - 1/2} \right)\left( {n + \nu  - 1/2} \right)} . \nonumber 
\end{eqnarray}

It follows that an arbitrary state can be found by the $n$-fold action of the $B^+$ to the ground state of $\psi _n$, i.e.

\begin{eqnarray}
\label{41}
\psi _n  = N_n \left( {B^ +  } \right)^n \psi _0 = c_n \left( { - \rho } \right)^{\left( \alpha  \right)} \omega _0 ^{i\rho } \Gamma \left( {\nu  + i\rho } \right)S_n \left( {\rho ^2 ;\alpha ,\nu ,1/2} \right), \\ 
 N_n ^{ - 1}  = b_1 b_2  \cdots b_n  = \left( {2\omega _0 } \right)^n \sqrt {n!\left( {\alpha  + \nu  + 1} \right)_n \left( {\alpha  + 1/2} \right)_n \left( {\nu  + 1/2} \right)_n } , \nonumber 
\end{eqnarray}
where $S_n \left( x^2; a,b,c \right)$ are the continuous dual Hahn polynomials. The generators $K^+$ and $K^-$ act on the wave functions $\psi _n$ in the following way:

$$
K^ -  \psi _n  = \kappa _n \psi _{n - 1} ,\;K^ +  \psi _n  = \kappa _{n + 1} \psi _{n + 1} ,\;\kappa _n  = \sqrt {n\left( {n + \alpha  + \nu  - 1} \right)} .
$$

\section{Conclusion}

In this paper in the framework of the finite-difference relativistic quantum mechanics we constructed the close Lie algebra of the $SU\left( 1, 1 \right)$ group for a relativistic model of the linear singular oscillator. In complete analogy with the non-relativistic problem, the relativistic problem under consideration is also exactly factorizable and this factorization allows us to obtain the exact finite-difference expression of the generators of $SU\left( 1, 1 \right)$ group. We show that eigenfunctions $\psi _n$ of a relativistic model of the linear singular oscillator form the basis of the irreducible representation $D^ +  \left( {\frac{{\alpha  + \nu }}{2}} \right)$ of the $SU\left( 1,1 \right)$ group.

\acknowledgments

One of the authors (E.I.J.) would like to acknowledge that this work is performed in the framework of the Fellowship 05-113-5406 under the INTAS-Azerbaijan YS Collaborative Call 2005.

% \appendix

% \section{}
% \label{}

\end{document}